\documentclass[10pt,twoside]{article}
\usepackage{amssymb}

\usepackage{graphicx}

\begin{document}

\title{Generalized Chaplygin gas with $\alpha = 0$\\
and the $\Lambda CDM$ cosmological model}
\author{J. C. Fabris\thanks{%
e-mail: \texttt{fabris@cce.ufes.br}}, S.V.B. Gon\c{c}alves\thanks{%
e-mail: \texttt{sergio@cce.ufes.br}} and R. de S\'a Ribeiro\thanks{%
e-mail: \texttt{rosilene-ribeiro@bol.com.br}} \\
\\
\mbox{\small Universidade Federal do Esp\'{\i}rito Santo, Departamento
de F\'{\i}sica}\\
\mbox{\small Av. Fernando Ferrari s/n - Campus de Goiabeiras, CEP
29060-900, Vit\'oria, Esp\'{\i}rito Santo, Brazil}}
\date{\today}
\maketitle

\begin{abstract}
The generalized Chaplygin gas model is characterized by the
equation of state $p = - \frac{A}{\rho^\alpha}$. It is generally
stated that the case $\alpha = 0$ is equivalent to a model with
cosmological constant and dust ($\Lambda CDM$). In this work we
show that, if this is true for the background equations, this is
not true for the perturbation equations. Hence, the mass spectrum
predicted for both models may differ.

\vspace{0.7cm}
\end{abstract}

PACS number(s): 98.80.Bp, 98.65.Dx \vspace{0.7cm}

The generalized Chaplygin gas ($GCG$) \cite{bento} is a recent
proposal in order to explain the observed acceleration of the
Universe \cite{riess,perlmutter}. It is an exotic fluid with
negative pressure whose equation of state is given by
\begin{equation}
\label{eoe}
p = - \frac{A}{\rho^\alpha} \quad ,
\end{equation}
with $0 \leq \alpha \leq 1$. This exotic fluid has been considered
as an alternative to quintessence \cite{steinhardt} and to the
cosmological constant \cite{sahni}, which are other serious
candidates to explain the accelerated expansion of the Universe.
Many observational constraints have been obtained for cosmological
models based on the $GCG$. One interesting aspect of such exotic
fluid is connected with a possible unification of dark matter and
dark energy through a simple fluid described by the equation of
state (\ref{eoe}).
\par
Some authors (see, for example, reference \cite{ioav}) claim that
the comparison with observation indicates that $\alpha$ is peaked
around zero. At same time, it is generally argued in the
literature that the case $\alpha = 0$ is equivalent to a $\Lambda
CDM$ model \cite{avelino,lima,ioav}. The aim of the present letter
is to show that, if this is true for the background solutions, it
is not true for the linearized equations. In this sense, in what
concerns the type Ia supernovae data both models leads to the same
results, as well as for the position of the acoustic peaks in
spectrum of the anisotropy of the cosmic microwave background
radiation. But we expect a disagreement concerning the predictions
for the mass spectrum.
\par
The case $\alpha = 0$ means that the pressure remains constant as
the Universe expands and the density decreases. Since, for this
case, $p = - A$, the equations of motion for an isotropic and
homogeneous Universe described by the flat
Friedmann-Robertson-Walker metric are:
\begin{eqnarray}
\label{em1}
3\biggr(\frac{\dot a}{a}\biggl)^2 &=& 8\pi G\rho \quad , \\
\label{em2}
2\frac{\ddot a}{a} + \biggr(\frac{\dot a}{a}\biggl)^2
&=& 8\pi GA
\quad , \\
\label{em3}
\dot\rho + 3\frac{\dot a}{a}(\rho - A) &=& 0 \quad .
\end{eqnarray}
These equations admit the solution,
\begin{eqnarray}
\label{s1}
a(t) &=& a_0\sinh^{2/3}(Dt) \quad , \\
\label{s2}
\rho(t) &=& A\coth^2(Dt) \quad ,
\end{eqnarray}
where $D = \sqrt{6\pi GA}$. The scale factor has exactly the same
behaviour as in a model with a dust fluid and a cosmological
constant. In fact, the integration of equation (\ref{em3}) leads
to the expression
\begin{equation}
\rho(t) = A + \rho_0a^{-3} \quad .
\end{equation}
Substituting this expression in equation (\ref{em1}), we obtain
the same relation for the scale factor of the $\Lambda CDM$ model.
Notice that the solution (\ref{s1}) display a dust behaviour when
$t \rightarrow 0$, and a cosmological constant behaviour when $t
\rightarrow \infty$.
\par
Since the equation of motion for the scale factor is the same as
in the $\Lambda CDM$, the luminosity-distance and angular diameter
relations will be same as in the $\Lambda CDM$ model. Hence, the
type Ia supernovae data and the location of the acoustic peaks in
the CMB anisotropy spectrum are the same in both cases. But, the
linearized equations around the background described above are
quite different, as it will be verified now.
\par
Introducing fluctuations around the background specified above,
and using the synchronous coordinate condition $h_{\mu0} = 0$, we
end up with the following coupled system of equations for the
$GCG$ with $\alpha = 0$:
\begin{eqnarray}
\ddot h + 2\frac{\dot a}{a}\dot h &=& 8\pi G\delta\rho \quad , \\
\dot\delta + 3\frac{\dot a}{a}\frac{A}{\rho}\delta &=&
\frac{1}{2}\biggr(1 - \frac{A}{\rho}\biggl)\dot h \quad ,
\end{eqnarray}
where
\begin{equation}
h = \frac{h_{kk}}{a^2} \quad \mbox{and} \quad \delta =
\frac{\delta\rho}{\rho} \quad .
\end{equation}
Using the background solutions and combining both equations, we
find the following equation for density contrast $\delta$:
\begin{equation}
\label{p1} \ddot\delta + \frac{4}{3}\biggr(\coth t + 3\tanh
t\biggl)\dot\delta + 2\biggr(\tanh^2t - \frac{1}{3}\coth^2t +
\frac{8}{3}\biggl)\delta = 0 \quad .
\end{equation}
In this equation, the time coordinate has been re-scaled: $Dt
\rightarrow t$. This equation for the density contrast can be
compared with that for the $\Lambda CDM$ model:
\begin{equation}
\label{p2} \ddot\delta + \frac{4}{3}\coth t\,\dot\delta -
\frac{2}{3}\coth^2t\,\delta = 0 \quad .
\end{equation}
\par
Equations (\ref{p1},\ref{p2}) exhibit the same solutions in the
``dust phase", when $t \rightarrow 0$: $\delta_+ \propto t^{2/3}$
and $\delta_- \propto t^{-1}$. But, in the ``cosmological constant
phase" the behaviour of (\ref{p1},\ref{p2}) are quite different.
In fact, in this limit, those equations simplify to
\begin{eqnarray}
\ddot\delta + \frac{16}{3}\dot\delta + \frac{20}{3}\delta &=& 0 \quad (GCG) \quad , \\
\ddot\delta + \frac{4}{3}\dot\delta - \frac{2}{3}\delta &=& 0
\quad (\Lambda CDM) \quad .
\end{eqnarray}
The equations admit solutions under the form of $\delta \propto
e^{nt}$. For the generalized Chaplygin gas model with $\alpha =
0$, $n_+ = - 5/3$ and $n_- = - 2$, while for the $\Lambda CDM$
model, $n_\pm = - 2/3 \pm \sqrt{10}/3$. Hence, in the $\Lambda
CDM$ phase there is a growing mode even during the ``cosmological
constant phase". On the other hand, for the generalized Chaplygin
gas model, there are only decreasing modes in the ``cosmological
constant phase". This agrees with the perturbative analysis made
for the Chaplygin gas with $\alpha = 1$ \cite{pasquier,patricia}.
\par
Equation (\ref{p1}) seems to admit no closed solutions. Hence, a
numerical integration must be performed. Frequently, in order to
fix the interval of integration, the functions are re-expressed in
terms of the redshift $z = -1 + a/a_0$, where $a_0$ is the scale
factor today. But, here, we will keep working with the re-scaled
cosmic time. The age of the Universe today will be fixed through
the deceleration parameter
\begin{equation}
q = - \frac{\ddot a a}{\dot a^2} \quad ,
\end{equation}
and the integration will be performed from the beginning of the
matter dominated phase, which occurs about $z \sim 4,000$.
Estimations of the deceleration parameter are consistent with a
value $q = - 0.77$ \cite{patricia2}. This leads to $t_0 \sim 1.$;
the initial time corresponding to $z = 4,000$ is $t_i \sim
10^{-5}$. At this initial moment, the density contrast will be
assumed to have a scale invariant spectrum with $\delta_i =
10^{-5}$.
\par
The result of the numerical integration for the $GCG$ model with
$\alpha = 0$ is displayed in figure $1$, while the result for the
$\Lambda CDM$ model is shown in figure $2$. The main difference is
due to the fact that, while for $\Lambda CDM$ the density contrast
for the dust fluid keeps growing even during the ``cosmological
constant phase", for the $GCG$ with $\alpha = 0$ it stops growing
and begins to decrease. The numerical integration confirms the
asymptotic analysis performed above.
\begin{figure}[t]
\begin{center}
\includegraphics[scale=0.80]{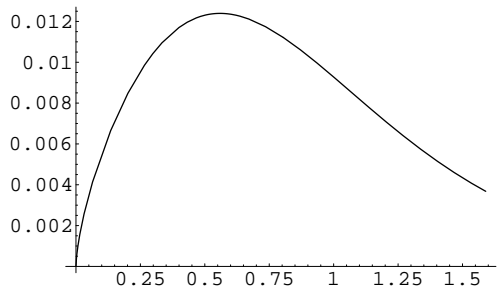}
\end{center}
\caption{{\protect\footnotesize Behaviour of the density contrast
for the $GCG$ model with $\alpha = 0$.}} \label{figA}
\end{figure}

\begin{figure}[t]
\begin{center}
\includegraphics[scale=0.80]{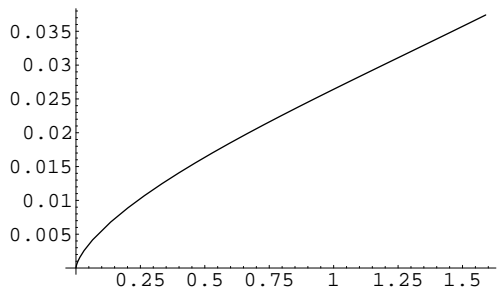}
\end{center}
\caption{{\protect\footnotesize Behaviour of the density contrast
for the $\Lambda CDM$ model.}} \label{figB}
\end{figure}
\par
It is curious to remark that the same behaviour for the background
may be found when the matter content of Universe is described by a
viscous fluid, if the bulk viscosity varies inversely with density
and if the spatial section is flat. In such a situation, the
background equations read
\begin{eqnarray}
3\biggr(\frac{\dot a}{a}\biggl)^2 &=& 8\pi G\rho \quad , \\
\dot\rho + 3\frac{\dot a}{a}(\rho + p^*) &=& 0 \quad ,
\end{eqnarray}
where $p^* = p - 3\frac{\dot a}{a}\xi(\rho)$. Fixing $p = 0$, and
$\xi(\rho) = \xi_0\rho^\nu$, the equations above become equivalent
to the $GCG$ model if $\nu + 1/2 = - \alpha$. The case $\alpha =
0$ is reproduced if $\nu = - 1/2$. Again this equivalence is valid
only for the background relations: at perturbative level, this
exotic viscous model is inequivalent both to the corresponding
$GCG$ model and to $\Lambda CDM$.
\par
The main consequence of this quite different behaviour concerns
the power spectrum for the mass distribution of the Universe.
Observational results on the mass power spectrum have been
obtained through the 2dFGRS program \cite{percival}, up to scale
of some hundreds of megaparsecs. A proper comparison with these
observational data implies, among other things, to evaluate the
transfer function for the $GCG$ with $\alpha = 0$. The evaluation
of the transfer function requires to consider a more detailed
model, including for example the radiative component of the matter
content of the Universe \cite{coles}. We hope to present a full
analysis of this problem in the future.
\newline
\vspace{0.5cm}
\newline
{\bf Acknowledgements:} We thank CNPq (Brazil) for partial
financial support to this work.

\end{document}